\documentclass[twocolumn]{jpsj3}
\usepackage{txfonts}

\title{Calculations of Branching Ratios for Radiative-Capture, One-Proton,
and Two-Neutron Channels in the Fusion Reaction $^{209}$Bi+$^{70}$Zn}

\author{Takatoshi Ichikawa and Akira Iwamoto $^{1}$}

\inst{Yukawa Institute for Theoretical Physics, Kyoto University, Kyoto 606-8502, Japan\\
$^{1}$Japan Atomic Energy Agency, Tokai-mura, Naka-gun, Ibaraki
319-1195, Japan\\}

\abst{We discuss the possibility of the non-one-neutron emission channels 
in the cold fusion reaction $^{70}$Zn + $^{209}$Bi to produce the element $Z=113$.  
For this purpose, we calculate the evaporation-residue cross sections of 
one-proton, radiative-capture, and two-neutron emissions
relative to the one-neutron emission in the reaction $^{70}$Zn + $^{209}$Bi.
To estimate
the upper bounds of those quantities, we vary model parameters
in the calculations, such as the level-density parameter and the height
of the fission barrier.
We conclude that the highest possibility is for the 2n reaction channel,
and its upper bounds are 2.4$\%$ and at most less
than 7.9\% with unrealistic parameter values, under the actual experimental
conditions of [J. Phys. Soc. Jpn. {\bf 73} (2004)
2593].}

\kword{nuclear reaction, superheavy element, cold fusion reaction, deexcitation, decay width}

\begin{document}
\maketitle

 \section{Introduction}
 Heavy-ion fusion reactions provide a powerful tool for synthesizing
 superheavy elements (SHE)~\cite{Hof00,Org07}.
 To verify that a new element has been created in such reactions, it is
 necessary to 
 identify which evaporation residue is 
 formed when the excited fusion product, in competition with
 nuclear fission, is de-excited by particle 
 or gamma-ray emissions to a certain stable state. 
 In this paper, we focus on
 the experiment performed by Morita {\it et al.} to produce
 the element 113~\cite{Mori04,Mori07}. The reaction  
 channel of this experiment for the observed two alpha chains was assigned as
 the one-neutron (1n) evaporation channel.
 We investigate here the sensitivity of this assignment to different
 model assumptions.

 The synthesized elements can be identified by measuring
 the alpha-decay chains to known nuclei.
 However, this is not always straightforward in the case of odd-odd
 evaporation residues, such as those in the experiments cited in refs.~3 and 4. 
 In such alpha-decay chains, one can expect not only
 ground-state-to-ground-state alpha decay but also the involvement of
 excited states.  
 Therefore, it is difficult to compare directly the observed alpha-decay
 energy and lifetime with the measured values of known products 
 in the decay chain.
 Uncertainties coming from experimental conditions may also lead
 to difficulties in identifications.
 Morita {\it et al.} have recently performed an 
 experiment on the formation of the nucleus $^{266}$Bh~\cite{Mori09}, which is the
 grand-grand 
 daughter of the nucleus $^{278}$113, and observed its decay modes.
 From the comparison of the alpha-decay chain of $^{266}$Bh with that of 
 $^{278}$113, they further confirmed the 1n reaction channel assignment. 

 In addition to these experimental studies, it is desirable to calculate
 the confidence level with which other reaction
 channels can be excluded from the reaction mechanism consideration.
 We therefore calculate the probabilities of the competing
 reaction channels relative to the 1n channel in
 standard theoretical models. Although the calculation is performed only for
 this special example, this approach can be applied to any other
 experiments for the SHE production based on  cold fusion reactions,
 offering a simple and practical estimator.  
  
 In practice, for the synthesis of the nucleus $^{278}$113 using the
 $^{209}$Bi($^{70}$Zn, n) reaction~\cite{Mori04}, Morita {\it
 et al.} excluded the following 
 possibilities: {\it (i) radiative capture process (zero-neutron
 evaporation channel) leading to the nucleus $^{279}$113, (ii)
 two-neutron evaporation channel leading to the nucleus $^{277}$113, and 
 (iii) one-proton evaporation channel leading to the nucleus
 $^{278}$112.} We will quantitatively estimate the probability of these
 decay channels by calculating the evaporation-residue (ER) cross
 section.

 For such estimations, it is difficult to calculate the non-1n reaction
 using the Monte-Carlo simulation for the de-excitation process because
 of its extremely low survival probability. In this respect, we employ
 the ``fusion-by-diffusion'' (FBD) model~\cite{Sw03,Sw05}, which consists of
 simple analytic expressions and is well optimized for various cold fusion
 reactions. We extend it to the non-1n reaction.
 One difficulty in the estimation is that the 
 de-excitation process is very sensitive to the level-density parameter
 and the height of the fission barrier.
 To reduce these uncertainties,
 we estimate worst-case scenarios for relative
 probabilities by varying substantially the level-density
 parameter and the height of the fission barrier.  
 We demonstrate below that the assignment given to the experiment cited in 
 ref.~3 is highly probable.
 
 This paper is organized as follows: In \S2, we describe details of models
 and parameters for numerical calculations. In \S3, we show
 the calculated results and discuss the probability of the radiative
 capture, two-neutron 
 evaporation, and one-proton evaporation channels. Finally, we
 summarize in \S4 the confidence level with which the evaporation residues of
 the experiments cited in refs.~3 and 4 can be assigned to the
 nucleus $^{278}$113.   
 
 \section{Model and Input Parameters}

 In this study, all calculations are based on the (FBD) model proposed by
 \'Swi\c{a}tecki {\it et al.}~\cite{Sw03,Sw05} 
 and its straightforward extensions,
 because their model and parameters have been well optimized for various 
 cold-fusion reactions and reproduced well the experimental
 data~\cite{Nel08-1,Dra08,Nel08-2,Gat08,Dra09,Fol09,Nel09}.
 The evaporation residue (ER) cross section $\sigma_{\rm ER}$ is given by
  \begin{equation}
   \sigma_{\rm ER}=\sigma_{\rm cap}\cdot P_{\rm diff}\cdot W_{\rm suv},
  \end{equation}
  where $\sigma_{\rm cap}$ is the capture cross section, $P_{\rm
  diff}$ is the probability that two touching nuclei can reach the
  compound state, which is descried as ``diffusion'' driven by thermal fluctuations of
  collective degrees of freedoms, and $W_{\rm suv}$ is the survival
  probability of the excited compound nucleus.
 
 One characteristic feature of this model is that the neutron decay width is
 calculated with the ``transition state'' method\cite{Sw05,Sw08,Ada10,Sw10}, rather
 than the traditional ``detailed-balance'' method~\cite{Van73}.
 They showed that the neutron evaporation width evaluated by the
 detailed-balance method with the temperature-dependent shell 
 correction energy is incorrectly suppressed near the fission
 threshold~\cite{Sw08}.
 We apply the transition state method to the proton evaporation channel
 as well.

 For the nuclear mass, the finite-range droplet model
 (FRDM95)~\cite{Mo95} and the Hartree-Fock-Bogoliubov model 
 (HFB02)~\cite{Go02} have been recommended in the reference input parameter 
 library (RIPL-2)~\cite{RIPL}.
 In the FBD model,
 the Thomas-Fermi model developed by Myers and \'Swi\c{a}tecki
 (MS96)~\cite{MS06} 
 was used. Among those models, we 
 chose FRDM95  as the main model in this  
 work, because not only the mass excess but also the shell
 correction energies and the fission barrier heights are 
 consistently provided.
 Although we use the parameters optimized for FRDM95 in all calculations, 
 we investigate the worst-case scenario at the different excitation
 energies estimated with the three mass models, which also causes
 uncertainty in this study, as will be discussed in \S2.1.  
 
 We calculate the level-density parameters $a$ with eq.~(A10) in
 ref.~7. 
 The uncertainty of the level density with the shell correction energy of
 FRDM95 was extensively investigated in a previous study cited in ref.~21.
 In RIPL-2, it was shown that the expected uncertainty for
 an unknown level-density parameter is about $6.5\%$. We
 thus vary the asymptotic level-density parameter by up to $10\%$ (see
 also Fig.~6.4 in ref.~21).
 
 \begin{table}
  \caption{Fission barrier heights $B_{\rm f}$ obtained from ref.~23 and
  one-neutron and one-proton separation energies $S_{\rm 1n}$ and
  $S_{\rm 1p}$ estimated with FRDM95~\cite{Mo95}.}
  \label{t1}
  \begin{center}
   \begin{tabular}{cccc}
    \hline
    \multicolumn{1}{c}{Nuclide} & \multicolumn{1}{c}{$B_{\rm f}$ (MeV)}
    & \multicolumn{1}{c}{$S_{1{\rm n}}$ (MeV)}&
    \multicolumn{1}{c}{$S_{1{\rm p}}$ (MeV)}\\ 
    \hline
    $^{279}$113 & 6.12 & 7.37 & 0.48\\
    $^{278}$113 & 6.06 & 6.32 & \\
    $^{277}$113 & 6.31 & 7.59 & \\
    $^{278}$112 & 5.99 & 7.14 & \\
    \hline
   \end{tabular}
  \end{center}
 \end{table}
 
 We use the height of the fission barrier tabulated in ref.~23 and the
 one-neutron and one-proton separation energies calculated with FRDM95.
 The values used in the calculation are tabulated in Table I.
 We also analyze the neutron separation energies calculated with the
 other two models and observe a small difference.
 For the uncertainty of the fission barrier height, the
 root-mean square deviation of the calculated results from the
 experimental data is shown to be 0.99 MeV in refs.~23 and 24; thus, we increase
 or decrease the
 height of the fission barrier by $\pm 1$ MeV.
 We also simultaneously increase or decrease the shell correction
 energy, because the fission barrier in the superheavy-mass region
 is mainly due to the shell correction energy at the ground state. 
 
 For calculations of $P_{\rm diff}$, we estimate the temperature
 at the saddle point with the fission barrier heights given in Table I. All
 the other calculations are the same as those in ref.~7. Because 
 the fission barrier heights differ from the original calculations in
 ref.~7, we need to fine-tune the 
 neck parameter $s$, which is only one adjustable parameter in the FBD
 model. We choose $s=2.1$ fm in order to fit the experimental data for
 the cold fusion reactions performed by RIKEN and obtain an overall fit
 for the experimental data, as shown in Fig.~1. In the figure, the solid
 and dashed lines respectively indicate the calculated results of the ER cross section
 with and without the target correction due to the target thickness
 under actual experimental conditions (as will be discussed in
 \S2.7). For comparison with the non-1n reaction channels, we fix all
 parameters for the 1n reaction channel.   

  \begin{figure}[t]
   \begin{center}
    \includegraphics[keepaspectratio,width=\linewidth]{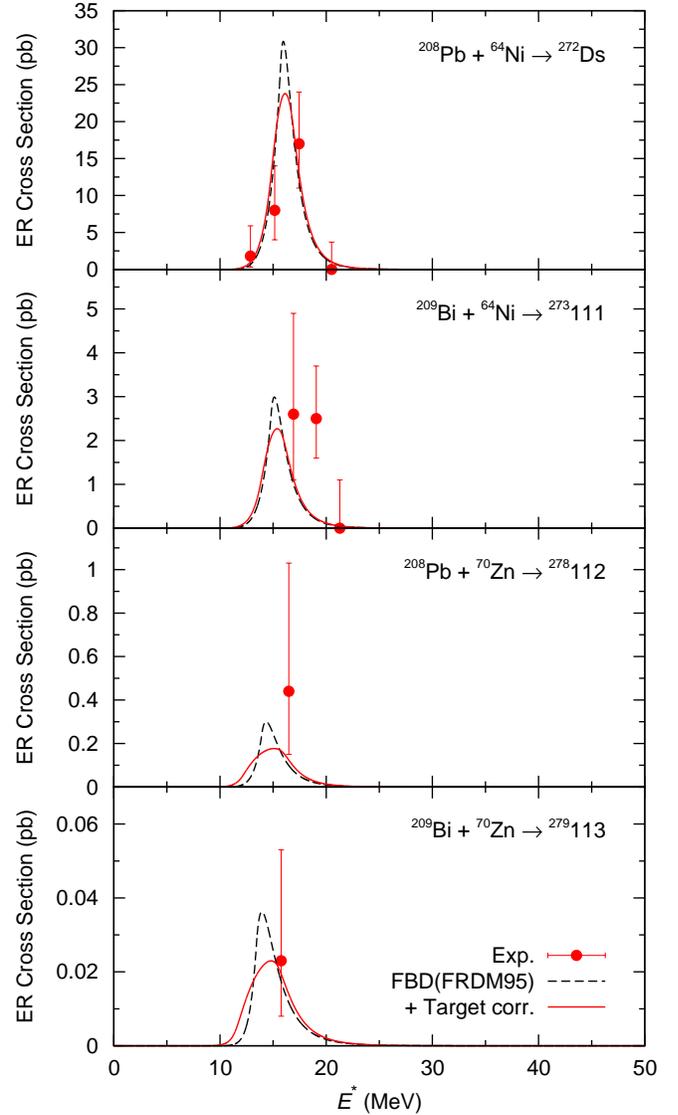}
   \end{center}
   \caption{(Color online) Evaporation residue cross sections for the 1n
   reaction channel in various cold fusion reactions calculated with 
   the FBD model.
   The solid and dashed lines
   respectively indicate the calculated results with and without the correction due to the
   target thickness. The solid circle indicates the experimental
   data obtained from refs.~25-28.
   }  
   \label{f5}
  \end{figure}
 
  \subsection{Range of excitation energies}
  
  One large uncertainty in estimations of the excitation energy of
  a formed compound nucleus is due to large differences between
  various mass models. For comparison, we calculate the excitation
  energies of the compound nucleus $^{279}$113 in the $^{209}$Bi +
  $^{70}$Zn reaction with FRDM95, MS96, and HFB02.
  For the mass excesses of $^{209}$Bi and $^{70}$Zn, we
  use the experimental masses obtained from Audi2003~\cite{AME}.
  Since the experimental
  incident beam energy is 261.4 MeV, the resultant excitation energies
  are 15.76, 14.11 and 13.61 MeV for FRDM95, MS96, and HFB02,
  respectively. 
  Note that in the experiment, the uncertainty of the incident beam
  energies is $\pm2$ MeV due to the target thickness~\cite{Mori04}.  
  We thus vary the excitation energies in the range from 11.61 to
  17.76 MeV in this study.
    
  \subsection{Proton decay width}

  We extend the transition state method to the proton decay. The
  particle decay width is then given as  
  \begin{equation}
   \Gamma_{p}=\frac{D}{2\pi}\int_{0}^{U^*-B_{\rm
    p}-V_c}\rho_F^{(p)}(U^*-B_{\rm p}-V_c-\epsilon)d\epsilon,
  \end{equation}
  where $B_{\rm p}$ is the binding energy of an evaporated proton, $V_c$ is
  the height of the Coulomb barrier, and $\rho_F$ is the 
  level density for the final state. The symbol $U^*$ is the effective excitation
  energy, which is defined by $U^* = E^*-\Delta$, where $E^*$ is the
  excitation energy and $\Delta$ is the pairing energy normalized to zero
  for odd-odd nuclei as given explicitly in APPENDIX B of ref.~7. The
  symbol $D$ is
  $1/\rho_I$, where $\rho_I$ is the level density for the initial state.
  
   In the calculation, the height of the Coulomb
   barrier, $V_c$, has a large uncertainty, due to the difficulty in
   its direct measurement.
   In the standard statical model, $V_c$ is estimated from the
   fusion barrier of its inverse reaction. The empirical fusion barrier
   $B_{\rm fus}$ is often taken as~\cite{Vaz84}  
   \begin{equation}
    B_{\rm fus}=\frac{1.44(Z-1)}{1.18(A-1)^{1/3}+3.928}.
   \end{equation}
   However, Moretto pointed out that, in a heavy-mass system, the
   height of the Coulomb barrier for a proton emission is lower than
   that of the inverse reaction due to the polarization of many 
   protons~\cite{Mor72}, and  Parker {\it et al.} proposed a new
   empirical formula extracted from the experimental
   systematics~\cite{Par91}, given by $B_{\rm fus}=0.106Z-0.90$.
   For the nucleus $^{279}$113, the heights of the Coulomb barriers are 
   13.87 and 11.08 MeV for the standard and new empirical
   formulas, respectively. In this paper, we use $V_c=11.08$ MeV to
   discuss the worst-case scenario.

  \subsection{$\gamma$-Ray decay width}

  For the gamma-decay width, we obtain equations and their parameters from
  ref.~33; however, for simplicity, we ignore the angular-momentum
  dependence and consider only the E1 transition. The gamma-decay width is thus
  given by
   \begin{equation}
    \Gamma_{\gamma}=\hbar\int_{0}^{U^*}w_{\gamma}(U^*,\epsilon_\gamma)d\epsilon_\gamma,
   \end{equation}
   where the transition probability $w_\gamma$ for the E1 gamma-ray
   emission is given by
   \begin{equation}
    w_{\gamma}(U^*,\epsilon_\gamma)=\frac{1}{\hbar}\frac{\rho_F(U^*-\epsilon_\gamma)}{\rho_I(U^*)}\epsilon_\gamma^3
     f_{\rm E1}(\epsilon_\gamma).
   \end{equation}
   The function $f_{\rm E1}$ is the gamma-strength function. Following
   ref.~33, $f_{\rm E1}$ was chosen as the following  
   Lorentzian function:  
   \begin{equation}
    f_{\rm E1} = 3.31 \times 10^{-6} {\rm MeV}^{-1}\frac{NZ}{A}\frac{\epsilon_\gamma\Gamma}{(E_0^2-\epsilon_\gamma^2)^2+\epsilon_\gamma^2\Gamma^2},
   \end{equation}
   where we consider the values of the width $\Gamma$ and the resonance
   energy $E_0$ calculated using the droplet model~\cite{Mye77}: $\Gamma=5$ MeV and
   $E_0$ is given by 
   \begin{equation}
    E_0=167.23A^{-1/3}(1.959+14.074A^{-1/3})^{-1/2}{\rm MeV}.
   \end{equation}
 
  \subsection{Survival probability for one-proton emission}
  
  We calculate the survival probability for the 1n reaction with the
  FBD model and extend it to the 1p and 2n reactions.
  In ref.~7, the survival probability for the 1n reaction,
  $W_{\rm suv}^{(1{\rm n})}$, is expressed as the product of the probability
  for the one-neutron evaporation and the survival probability against
  the second-chance fission or the second-chance neutron emission, $W_<^{(\rm
  2nd)}$, such that we extend it to the survival probability for the 1p
  reaction, $W_{\rm suv}^{(\rm 1p)}$, given by  
   \begin{equation}
    W_{\rm suv}^{(1{\rm p})}(E^*)=\frac{\Gamma_{\rm p}(E^*)}{\Gamma_{\rm Tot}(E^*)}W_<^{(\rm 2nd)}(E^*),
   \end{equation}
   where $\Gamma_{\rm Tot}$ is the total decay width defined by
   $\Gamma_{\rm Tot}=\Gamma_{\rm n}+\Gamma_{\rm f}+\Gamma_{\rm p}+\Gamma_\gamma$.
   By assuming a standard proton evaporation spectrum proportional to
   $k\exp(-k/T)$, where $k$ is the neutron kinetic energy and $T$ is the
   temperature of the transition state for proton emission, $W_<^{(\rm
   2nd)}$ is given by
   \begin{equation}
    W_<^{\rm (2nd)}=
     \begin{cases}
      (1+\frac{K}{T})\exp(-\frac{K}{T}) & K \geqq 0,\\
      1 & K < 0,
     \end{cases}
   \end{equation}
   where $K=U^*-B_{\rm p}-V_C-E_{\rm th}$ and $E_{\rm th}$ is the threshold
   energy for the second-chance fission or second-chance neutron
   emission whichever is lower.  

  \subsection{Survival probability for two-neutron emission}
   
   For the 2n reaction, we assume that the survival probability is
   given by the product of $\Gamma_{\rm n}^{(1{\rm n})}/\Gamma_{\rm
   Tot}^{(1{\rm n})}$ for the first neutron evaporation, $\Gamma_{\rm
   n}^{(2{\rm n})}/\Gamma_{\rm Tot}^{(2n)}$ for the 
   second neutron evaporation, and the survival probability against the
   third-chance fission or neutron emission. 
   Since the second neutron evaporation width depends on the kinetic energy of
   the first emitted neutron, we approximate the probability of the
   second neutron emission using the ensemble average in the first neutron
   emission spectrum. $W_{\rm suv}^{(2{\rm n})}$ is thus given as  
   \begin{gather}
    W_{\rm suv}^{(2{\rm n})}(E^*)\sim\frac{\Gamma_{\rm n}^{(1{\rm n})}(E^*)}{\Gamma_{\rm Tot}^{(1{\rm n})}(E^*)}
     \left<\frac{\Gamma_{\rm n}^{(2{\rm n})}(E^*-B_{\rm n}-\epsilon)}{\Gamma_{\rm Tot}^{(2{\rm n})}(E^*-B_{\rm n}-\epsilon)}
     W_<^{(\rm 3rd)}(E^*-B_{\rm n}-\epsilon)\right>_\epsilon,\\
    \sim\frac{\Gamma_{\rm n}^{(1{\rm n})}(E^*)}{\Gamma_{\rm Tot}^{(1{\rm n})}(E^*)}
     \frac{\Gamma_{\rm n}^{(2{\rm n})}(E^*-B_{\rm n}-<\epsilon>)}{\Gamma_{\rm Tot}^{(2{\rm n})}(E^*-B_{\rm n}-<\epsilon>)}
     W_<^{(\rm 3rd)}(E^*-B_{\rm n}-<\epsilon>),
   \end{gather}
   where $W_<^{\rm (3rd)}$ is the probability for the third-chance fission or
   neutron emission, such as that given in eqs.~8 and 9,
   $<\dots>$ is the ensemble average, and $\epsilon$ 
   is the kinetic energy of the second emitted neutron.
   The quantity $<\epsilon>$ can be obtained as $<\epsilon>=2T$, assuming the
   Boltzman distribution for the neutron evaporation spectrum.
   
  \subsection{Survival probability for radiative-capture reaction}
   In an analogy with the calculation of the 1n reaction, 
   we calculate the survival probability $W_{\rm suv}^{(0{\rm n})}$ for
   the radiative-capture 
   reaction as the product of the probability
   of gamma emission and  survival probability against the second-chance
   fission or neutron emission after a gamma-ray emission, 
   $W_<^{(0{\rm n})}$, given by  
   \begin{equation}
    W_{\rm suv}^{(0{\rm n})}(E^*)=\frac{\Gamma_\gamma(E^*)}{{\Gamma_{\rm Tot}}(E^*)}W_<^{(0{\rm n})}(E^*).
   \end{equation}
   We can calculate $W_<^{(0{\rm n})}$ with the probability below the                
   fission or neutron emission threshold energy after a gamma-ray
   emission using the 
   gamma-ray spectrum of eq.~5. $W_<^{(0{\rm n})}$ is thus given by
   \begin{equation}
    W_<^{(0{\rm n})}=\frac{\int^{U^*}_{E_{\rm
     th}}w_{\gamma}(U^*,\epsilon_\gamma)d\epsilon_\gamma}{\int_{0}^{U^*}w_{\gamma}(U^*,\epsilon_\gamma)d\epsilon_\gamma},   
   \end{equation}
   where $E_{\rm th}$ is the threshold energy for the second-chance
   fission or neutron evaporation, whichever is lower.

    \begin{figure}[t]
     \begin{center}
      \includegraphics[keepaspectratio,width=\linewidth]{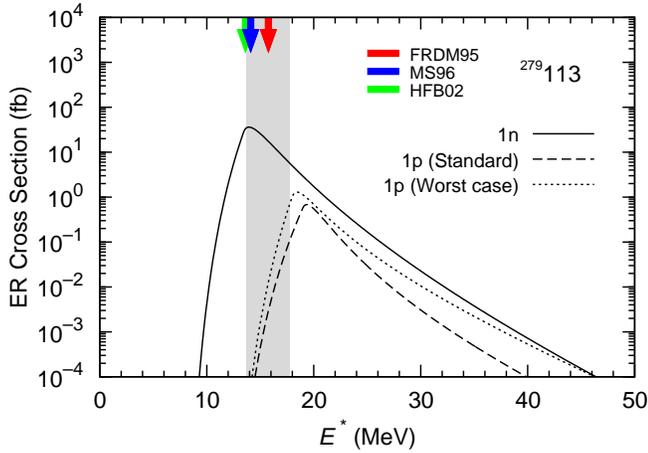}
     \end{center}
     \caption{(Color online) Evaporation residue cross sections for
     one-proton and one-neutron emissions versus excitation
     energy for the compound nucleus $^{279}$113. The solid and dashed
     lines respectively indicate the calculated results for the 1n and
     1p reaction channels with 
     the standard level-density parameter and fission barrier
     height. The dotted line shows the calculated result of the 
     1p reaction channel for the worst-case scenario with $a_{\rm
     p}\times 1.1$.  
     The dark-gray, black, and gray arrows indicate the excitation
     energies corresponding to the experimental incident energies
     calculated using FRDM95, MS06, and HFB02, respectively.
     The light-gray area denotes the incident-energy distribution for
     the worst-case scenario.}
     \label{f1}
    \end{figure}

  \subsection{Expected value relative to 1n reaction channel}
  To discuss the possibility of the non-1n reaction channel under actual
  experimental conditions, we calculate the expected value relative to 
  the 1n reaction channel, taking into account the effect of the target
  thickness, given by 
   \begin{equation}
    P=\frac{\int_{E_{\rm min}^*}^{E_{\rm
     max}^*}\sigma_{\rm ER}(E^*)dE^*}{\int_{E_{\rm min}^*}^{E_{\rm
     max}^*}\sigma_{\rm ER}^{(1{\rm n})}(E^*)dE^*},
   \end{equation}
   where $E^*_{\rm max}$ and $E^*_{\rm min}$ are the upper and lower
   bounds of the incident-energy distribution due to the target
   thickness, respectively.
   One advantage of this equation is that we can partially reduce the
   effects of uncertainties in $\sigma_{\rm cap}$ and $P_{\rm diff}$
   defined in eq.~1. 
  
 \section{Result and Discussion}
  \subsection{Probability for one-proton evaporation channel}

   To discuss the probability for the $1p$ channel relative to the
   1n channel, we calculate $\sigma_{\rm ER}^{(\rm 1p)}$ and compare it
   with $\sigma_{\rm ER}^{(\rm 1n)}$.
   Figure 2 shows the calculated result versus excitation energy for
   the nucleus $^{279}$113.    
   The solid and dashed lines indicate the results for the 1n and
   1p reaction channels with the standard
   level-density parameter and fission barrier height, respectively.
   For the standard parameter set, the excitation energy of the
   worst-case scenario is estimated to be 17.76 MeV with FRDM95, and the
   resultant $E_{\rm max}^*$ and $E_{\rm
   min}^*$ are 19.76 and 15.76 MeV, respectively. The incident-energy
   distribution is denoted by the
   light-gray area in the figure. In this case, we determine the expected
   value of the 1p reaction 
   channel relative to the 1n reaction channel, $P^{(\rm 1p/1n)}$, to be
   $8.51\times 10^{-4}$, which is a negligibly small probability for
   one-proton evaporation.   
   
   To determine the uncertainty originating from the level-density
   parameter and fission barrier height, we calculate $\sigma_{\rm
   ER}^{\rm (1p)}$ with $a_{\rm p}\times 1.1$ and 
   a decrease in fission barrier height of $^{278}112$ by 1 MeV (We also
   change consistently the shell correction energy at the
   ground state). 
   The calculated results are indicated by the dotted line in Fig.~2.
   Even in this case, the expected value of one-proton
   emission relative to one-neutron emission is less than
   $4.97\times 10^{-3}$. The branching ratio
   of  one-proton emission is thus negligibly small.

 \subsection{Probability for radiative-capture channel}

   \begin{figure}[t]
    \begin{center}
     \includegraphics[keepaspectratio,width=\linewidth]{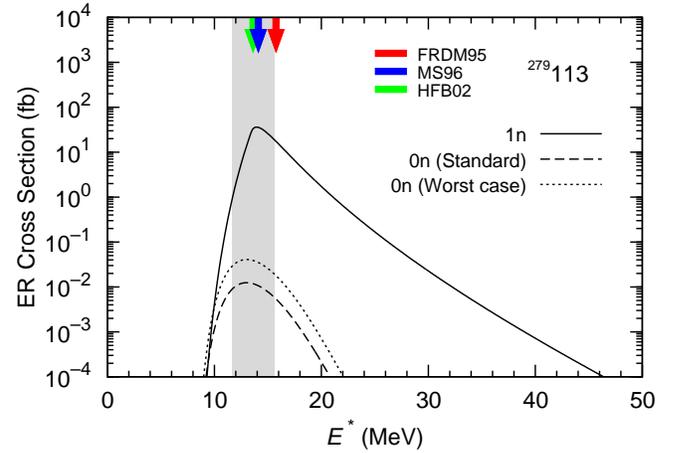}
    \end{center}
    \caption{(Color online) Evaporation residue cross sections for 
    the radiative-capture and one-neutron emission channels versus
    excitation energies of the compound nucleus $^{279}113$. The solid
    and dashed lines respectively indicate the calculated results for the 1n and
    0n reaction channels with standard parameters. The dotted line shows the
    results with $a_\gamma\times 1.1$ and a decrease in fission barrier height.
    The arrows and the gray area are the same as those in
    Fig.~1.} 
    \label{f4}
   \end{figure}
 
   Here, we discuss the probability of the radiative-capture and 1n
   reaction channels. 
   We calculate $\sigma_{\rm ER}^{(\rm 0n)}$ and compare it with
   $\sigma_{\rm ER}^{(\rm 0n)}$.
   Figure 3 shows the calculated result. The solid and dashed lines
   indicate 
   the calculated results for the 1n and 0n reaction channels with the
   standard level-density parameter and the fission barrier height,
   respectively. The excitation energy in the worst-case scenario is
   estimated to be 13.61 MeV with HFB02, and the resultant $E_{\rm max}$ and
   $E_{\rm min}$ are 15.61 and 11.61 MeV, respectively. For this case, the
   expected value of 
   the 0n reaction channel relative to the 1n reaction channel, $P^{(\rm
   0n/1n)}$, is $5.14\times 10^{-4}$.

   We also increase the level-density
   parameter by $10\%$ to estimate the worst-case scenario. For the
   radiative-capture reaction, we do not change the height of the 
   fission barrier of $^{279}113$ for the calculation of
   the second-chance fission, because
   such changes affect the calculated result for the 1n reaction
   channel. The calculated result 
   is indicated by the dotted line in the figure. For this worst-case
   scenario, $P^{(\rm 0n/1n)}$ is less than $1.70\times 10^{-3}$, which
   is so small that the radiative-capture channel can be excluded.
   
   \subsection{Probability for two-neutron evaporation channel}

  \begin{figure}[t]
   \begin{center}
    \includegraphics[keepaspectratio,width=\linewidth]{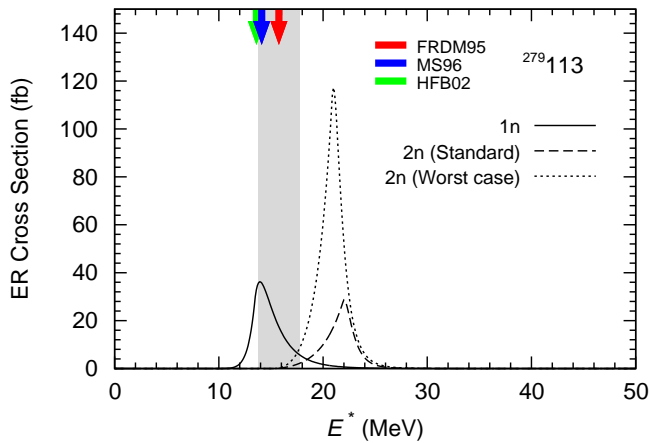}
   \end{center}
   \caption{(Color online) Evaporation residue cross sections for the 2n
   and 1n reactions versus excitation energy of the compound nucleus 
   $^{279}113$. The solid and dashed lines respectively indicate the
   calculated results for the 1n and 2n reaction channels with 
   standard parameters.
   The dashed line shows the result obtained with the level-density parameter
   $a_{n}\times 1.1$ for the 2n reaction and a decrease in 
   fission barrier height by 1 MeV (We also change the corresponding
   shell correction energy at the ground state). 
   The arrows and the light-gray area are the same as those in Fig.~1.}  
   \label{f5}
  \end{figure}
   
   To study the probability for the two-neutron evaporation
   channel, we calculate $\sigma_{\rm ER}^{(2{\rm n})}$.
   To estimate the level-density parameter at  
   the saddle point, we use the elongation of the major semi axis at the
   saddle configuration relative to the spherical shape, which is 2.5 
   fm, the recommended value for the deformed nucleus at the
   ground state in ref.~7.
   We also determine 
   the sensitivity of the calculation to this parameter and observe that
   it is negligibly small.
   
   We first calculate $\sigma_{\rm ER}^{(\rm 2n)}$ with the standard
   parameter set and show the results in Fig. 4.
   The solid and dashed lines respectively indicate the calculated results for the 1n and
   2n reaction channels with standard parameters. 
   In the figure, we see that the probability of the 2n reaction cannot
   be excluded only for the excitation energy estimated with FRDM95. The
   other two estimations are well below the threshold 
   energy for the 2n reaction, even if we take into account the
   effect of the target thickness. The expected value of the 2n reaction
   channel relative to the 1n reaction channel, $P^{(\rm 2n/1n)}$, is
   $2.44\times 10^{-2}$; thus, we can safely neglect the possibility of the 2n
   reaction channel.   

   We next discuss the uncertainty originating from the level-density
   parameter and the fission barrier height. We increase 
   $a_{\rm n}$ for the 2n reaction channel by $10\%$ and decrease
   the height of the fission barrier of $^{277}113$ and the shell
   correction energy by 1 MeV. The calculated result is indicated by the
   dotted line in Fig. 4. In the figure, we can see an unrealistic
   enhancement of the 2n reaction channel, compared with the fact that in all the experiments of
   the cold fusion reactions, the observed peak values of the ER cross section for
   the 1n reaction are higher than or comparable to that of the
   2n reaction~\cite{Hof00}. 
   For this case, $P^{(\rm 2n/1n)}=7.91\times 10^{-2}$, which is still a
   fairly small probability for the 2n reaction despite such unrealistic
   parameter values.
   
  \subsection{Probability for total non-1n evaporation channel}
   Finally, we calculate the total non-1n branching ratio, $P^{(\rm
   non-1n/1n)}$, using the sum of the probabilities for the proton,
   radiative-capture, and 2n reaction channels. For the standard
   parameter set, we obtain $P^{(\rm non-1n/1n)}=2.55\times 10^{-2}$,
   indicating that the dominant component of the non-1n channels is the 2n
   evaporation channel.

   For the worst-case scenario, we simultaneously
   use all the parameter
   sets for each channel discussed in the previous sections.
   Although we independently change the model parameters for each
   channel in discussing the worst-case scenario, their effects on the
   other channels are very small, because we only change
   the parameters for the different residual compound nuclei at the final state
   and fix those for the 1n evaporation channel in order
   to reproduce the experimental data.
   With all the parameters for the worst-case scenario, we calculate the
   expected values at the excitation energies estimated 
   with FRDM95 and obtain $P^{\rm (1p/1n)}=4.97\times 10^{-3}$, $P^{\rm
   (0n/1n)}=9.94\times 10^{-4}$, and $P^{\rm (2n/1n)}=7.92\times
   10^{-2}$. The total non-1n branching ratio is then $P^{(\rm
   non-1n/1n)}=8.51\times 10^{-2}$, which is still a fairly small
   probability. 

 \section{Conclusions}

 We have discussed the possibilities of proton emission, 
 radiative-capture, and two-neutron emission channels in the de-excitation
 of the compound nucleus $^{279}$113 produced in the $^{209}$Bi +
 $^{70}$Zn reaction reported in ref.~3. 
 To this end, we extend the FBD model to those channels.
 We vary the ratio of the level-density parameter to the
 height of the fission barrier to determine its effect on calculations
 and determine the upper bounds of those channels.

 To discuss the branching ratio, we calculate the expected value of the
 other reaction channel relative to the 1n reaction channel under actual
 experimental conditions.  
 For the one-proton emission channel, we obtain $P^{\rm
 (1p/1n)}=8.51\times 10^{-4}$ and at most 
 $4.97\times 10^{-3}$, indicating that the probability of the one-proton
 emission channel can be ignored.
 For the radiative-capture channel, we obtain $P^{\rm
 (0n/1n)}=5.14\times 10^{-4}$ and at most $1.70\times 10^{-3}$.
 We thus consider that the possibility of the radiative-capture channel can
 be excluded.
 For the two-neutron emission channel, we obtain $P^{\rm
 (2n/1n)}=2.44\times 10^{-2}$ and at most $7.91\times 10^{-2}$, which is
 a fairly small probability despite the unrealistic parameter values.  
 For the total non-1n branching ratio, we obtain $P^{(\rm
 non-1n/1n)}=2.55\times 10^{-2}$ and at most $8.51\times 10^{-2}$.
 We conclude that the 2n reaction is the main branch in the
 non-1n reaction channels.
 
 For the estimations of the worst-case scenario, the most sensitive
 parameter is the nuclear mass. The development of a mass model would
 largely improve the determination of the optimum bombarding energy
 in synthesizing new elements.

 \acknowledgments
 We dedicated this paper to the memory of W{\l}adek \'Swi\c{a}tecki, whose
 works on fission and superheavy elements were an inspiration to us
 throughout our careers.
 We would like to thank P.~M\"oller for valuable discussions.

\end{document}